# MashUp : Scaling TCAM-based IP Lookup to Larger Databases by Tiling Trees

Victor Rios
victorrios0343@gmail.com
George Varghese
UCLA
varghese@cs.ucla.edu

## ABSTRACT

Ternary content addressable memories (TCAMs) are commonly used to implement IP lookup, but suffer from high power and area costs. Thus TCAM included in modern chips is limited and can support moderately large datasets in data centers and enterprises, but fails to scale to backbone WAN databases of millions of prefixes. IPv6 deployment also makes it harder to deploy TCAMs because of the larger prefixes used in the 128-bit address space.

While the combination of algorithmic techniques and TCAM has been proposed before for *reducing power consumption* or *update costs* (e.g., CoolCAM [32] and TreeCAM [28]), we focus on *reducing TCAM bits* using a scheme we call MashUp that can easily be implemented in modern reconfigurable pipeline chips such as Tofino-3. MashUp uses a new technique, *tiling trees*, which takes into account TCAM grain (tile) sizes. When applied to a publicly available IPv6 dataset using Tofino-3 TCAM grain sizes (44 by 512), there was a 2X reduction in TCAM required. Further, if we mix TCAM and SRAM using a new technique we call *node hybridization*, MashUp decreases TCAM bits by 4.5X for IPv6, and by 7.5X for IPv4, allowing wide area databases of 900,000 prefixes to be supported by Tofino-3 and similar chips.

## 1 INTRODUCTION

An issue that has plagued the Internet since the dawn of Classless Internet Domain Routing (CIDR) is performing IP prefix lookup at wire speeds. The worldwide Internet of over a billion nodes is compressed into around 900,000 IPv4 prefixes today at the cost of Internet routers doing a longest matching prefix operation on every packet instead of a simple exact match. While this can be done by walking a tree that branches on address bits (trie), algorithmic solutions can be slow.

A simpler solution is a *Ternary Content Addressable Memory* or TCAM which employs hardware parallelism to match a destination address to all the stored prefixes in one clock cycle, and uses a priority encoder to pick the longest match. The TCAM must be ternary to allow storing prefixes because bits can either be 1, 0, or the don't care bit * . To implement a TCAM, 16 transistors are required per bit (compared to 6 for RAM) for masking and comparison, leading to increased area costs; further, most transistors must be active during the search, leading to increased power costs [1].

The costs are not so high that TCAM is unusable, but for situations where chip real-estate or power usage are critical, as in *wide area* routers, currently TCAM-based lookups are not considered a scalable option. Excessive TCAM can consume 10-100 watts[1] in power-constrained routers. Thus, for instance, for the wide area Arista 7500R3/7800R3 and 7280R3 routers that scale to more than 2.5 million routes, their documentation [6] states that "internally FlexRoute uses an *algorithmic* approach to performing lookups".

In this paper, we ask: *can limited TCAM memory be used by wide area routers for IP lookup, and for large IPv6 deployments in data centers?* First, we examine three relevant trends.

**Trend 1, TCAM enabled reconfigurable chips:** Chips, such as Barefoot's Tofino-3, have emerged that have a fairly large amount of TCAM. Intel's Aurora 710, based on Intel Tofino 3.2T switching silicon, claims to allow data centers to increase the IP routing table size from 300K to 1.2M, allowing them "to build even bigger networks and address many more servers" [8]. Further, the TCAM is distributed among a set of physical stages that can be programmed using P4. While this seems to offer a positive answer to our question, we note two other countervailing trends.

**Trend 2, IPv6 Deployment:** In 1988, as a response to IPv4 address depletion, the IETF created 128-bit IPv6. IPv6 initially languished because of the widespread use of Network Address Translation (NAT). However, the proliferation of mobile and Internet of Things (IoT) devices led to rapid deployment of IPv6. Statistics [10] show the IPv6 availability of Google users at over 30%. The IPv6 launch web site [26] reports Carrier networks and ISPs are leading the charge. For example, T-Mobile has more than 90% of its traffic going over IPv6. IPv6 has even reached clouds and data centers such as Azure [9]. Thus even if the Aurora 710 can support 1.2M *IPv4 prefixes*, it will be hard pressed to support as many *IPv6 prefixes* where the address space requires 4 times as many bits.

**Trend 3, Prefix Table Growth:** The prefix table continues to grow, even for IPv4. In the past year[2], IPv4 prefixes have grown at 6% and IPv6 at 33%. Based on current trends, the APNIC blog [2] estimates IPv6 database size to reach 128k by the end of 2022, and the IPv4 database to reach 1 million prefixes by the end of 2023. Similarly, prefix tables continue to grow in data centers [8].

**Solution**: To answer our question, we will exploit Trend 1 (the increased TCAM available in reconfigurable chips) to tackle Trends 2 and 3. We use a "mash up" of algorithmic and hardware techniques via a tree of TCAM and SRAM blocks. More specifically, we show how a unibit trie representing the database can be compactly tiled (Figure 1) while respecting TCAM grain sizes and yet minimizing pointers and wasted space within TCAM blocks. As a consequence, MashUp can extend the reach of chips like Tofino-3 to scale to the backbone IPv4 database, or to much large data centers than is possible using today's solution of a single logical TCAM.

A quick preview of the solution is as follows. First, TCAMs come in units with a *grain size* of $W$ (width) by $D$ (depth) like memory pages. The Tofino-3, for example, has $W = 44$ and $D = 512$. Each unit TCAM or block in Tofino-3 can do a longest match on up to 512 prefixes of up to 44 bits (any of which can be *) in a single clock cycle. For example, to fit the current IPv6 database size of 150,000 prefixes of up to 64 bits, the current approach is to stitch together 2 TCAM blocks horizontally ($64 < 88$) and 300 of these pairs vertically ($150{,}00/512 \approx 300$).

By contrast, MashUp builds a tree of TCAM blocks (as seen in Figure 1) starting with a root which in this picture fits (is tiled) into a TCAM block. This breaks up the bit-by-bit branching tree (trie) of prefixes into subtrees which we can recursively tile. To pick the tree level at which to define the root TCAM, observe that compared to a single logical TCAM the tree has pointers, which represent "overhead".

The first insight (Figure 1) is to cut the trie at a lean level $L(b)$, where the number of downstream pointers is small, defined as $b$% of the total database size, where $b$ is a small rational number. For example, for the IPv6 database, Table 2 shows that at level 20, the pointer overhead ($b$) is only 0.41% but rises to 6.84% at level 32.

The second insight is to reduce the wasted space of tiled subtrees at the second level by packing in units of $D$ subtrees. Because the remaining prefix bits may be repeat across different subtrees, this requires adding a disambiguating tag (Figure 1) of $\log_2 D$ bits but it ensures that any wasted space in the last block is amortized over $D$ subtrees.

The final insight is to do a "currency exchange" as it were, and replace "nearly full" subtrees by SRAM pages in a process called RAM hybridization (see bottom right of Figure 1). Since SRAM cannot do variable length matching, this must be remedied by "expanding" variable length prefixes to the maximum length in the subtree. Hybridization allows currency arbitrage because SRAM pages are cheaper and more plentiful than TCAM blocks (3 to 1 in Tofino-3). Figure 1 shows a 2-stage tree but we have found that our best results for IPv4 of 900k prefixes (§6) comes by using four stages in strides of 16-4-4-8, which reduces TCAM bits by 7X at the cost of using around 1000 SRAM pages.

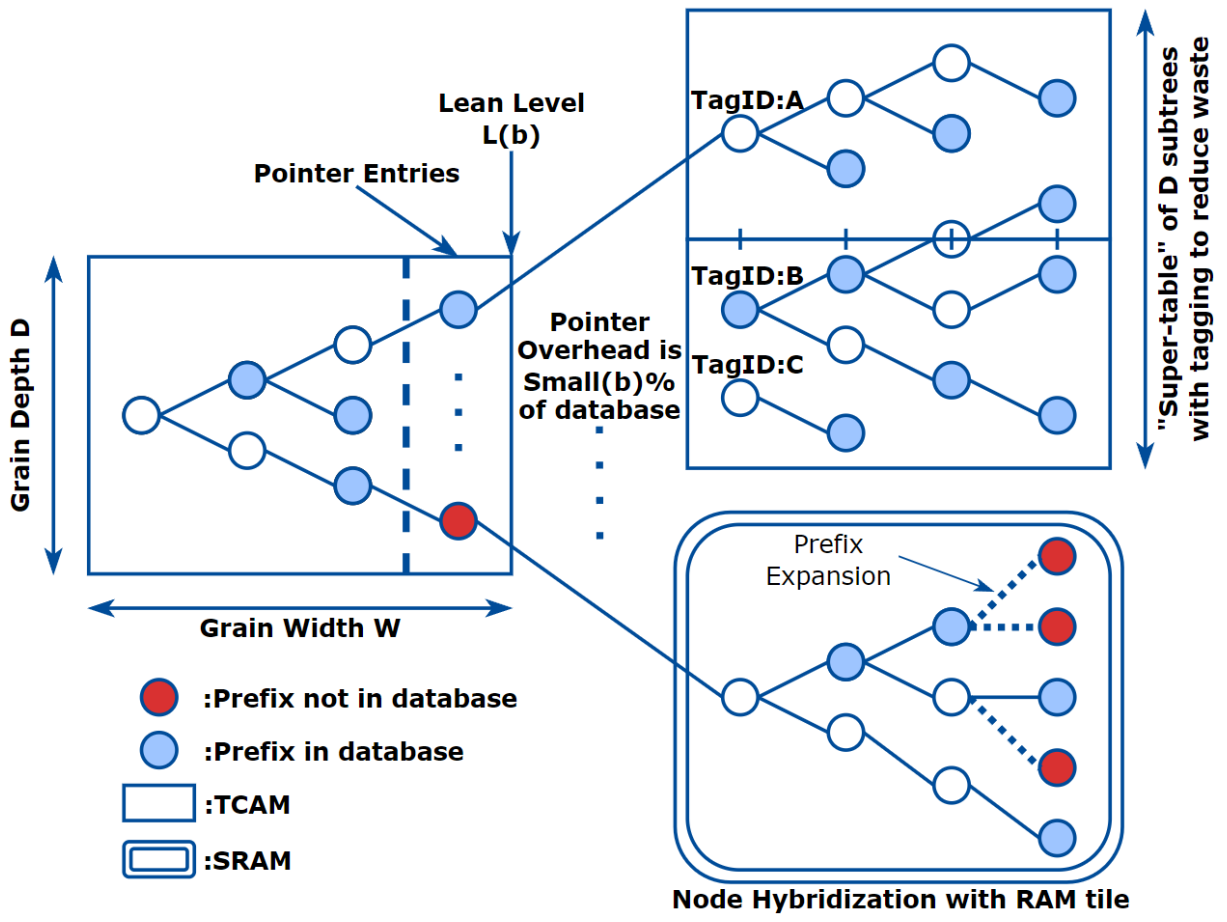

**Figure 1: Tiling a tree with MashUp**

The contributions of this paper are as follows:

*1. Techniques for Tiling Trees:* While trees of TCAMs have been used before, they have been used to reduce power (*e.g.* CoolCAMs [32]) and improve update costs (*e.g.* TreeCAM [28]). While we may share these benefits as a side effect, our main focus is increasing the size of supported IP prefix databases. This leads to different optimizations and a new set of *tree tiling* techniques via ideas like lean levels, tag aggregation, and per node hybridization (Figure 1).

*2. Analyzing the impact of TCAM Grain Sizes:* Unlike earlier papers that explore fundamental changes to the internal TCAM architecture, we focus on using existing TCAM resources in novel ways. We focus on adapting the tree parameters to the grain size of the underlying TCAMs *and* the characteristics of the database, in ways not seen in older work (*e.g.* a Tiling Theorem, Theorem 5.5.)

*3. Results:* Using modern TCAM grain sizes (*e.g.* 44 by 512 in Tofino-3), we show that we can support roughly 4.5X the number of IPv6 prefixes supported by a single logical TCAM. We also show how to shoehorn the current wide area IPv4 database of 900,000 prefixes in Tofino-3 whereas only 1/5-th of the database can be supported using the solution of a single logical TCAM (§6).

The rest of the paper is organized as follows. We review relevant background information in §2. The problem is described in §3 and a small example highlighting how TCAM can be inefficiently allocated is shown. In §4, the solution and methods for implementation are described. §5 presents a theoretical model for optimal savings as well as achieving maximum savings given database and TCAM characteristics. Experimental results and analysis obtained from using real BGP tables are presented in §6. Finally, we mention related work in §7 and conclude with §8. This work does not raise any ethical issues

| *Entry No.* | *Prefix(Ternary)* | *Output Port* |
|---|---|---|
| 1 | 1***** | A |
| 2 | 1000** | B |
| 3 | 10001* | C |
| 4 | 10010* | D |
| 5 | 100110 | E |
| 6 | 100111 | F |

**Table 1: Example Routing Table**

## 2 BACKGROUND

In this section, we review technologies and ideas that enable the implementation of MashUp.

**TCAMs:** Ternary Content Addressable Memories (TCAMs) are a form of associative memory implemented in hardware. During use, a search key is provided and is compared in parallel against an entire table of stored keys in a single clock cycle. Due to the ternary nature of stored bits, it is possible for there to be more than one match, in which case a priority encoder can be used to choose the first match. Longest matching prefix is implemented by keeping prefixes in the TCAM sorted by length. Using the returned result, a decoder is used to index into RAM in order to decide what action should be taken [21].

**Reconfigurable Match Tables (RMTs)**: The RMT pipelined architecture proposed by Forwarding Metamorphosis [4] consists of: a parser that produces a packet header vector, a series of logical stages that perform Match-Action operations, a recombination block to reattach headers to packets, and configurable output queues. Logical stages are mapped to one, multiple, and/or fractions of physical stages. The RMT model provides a natural setting for MashUp because every level of the tree of TCAMs can be placed in a logical stage(Figure 5b).

**Algorithmic Techniques:** MashUp is motivated by the trie data structure. A trie can be used to search any hierarchical namespace, such as the IP address space. Figure 2 shows a unibit trie for the database in Table 1. A significant disadvantage of using a unibit trie is that, in the worst-case, the number of memory accesses required is equal to the length of the address space (32 for IPv4 and 128 for IPv6).

Multibit tries offer a faster alternative to unibit tries. In a multibit trie, instead of looking at a single bit, multiple bits are used to find the next node. Figure 3 shows a multibit trie for Table 1 (the tree uses a stride list of 3-3, going 3 bits at the root and 3 bits at the next level). Decreasing the height of the tree reduces the number of memory access required to search the tree. The drawback of using multibit tries, however, is that *prefix expansion* is required which increases storage.

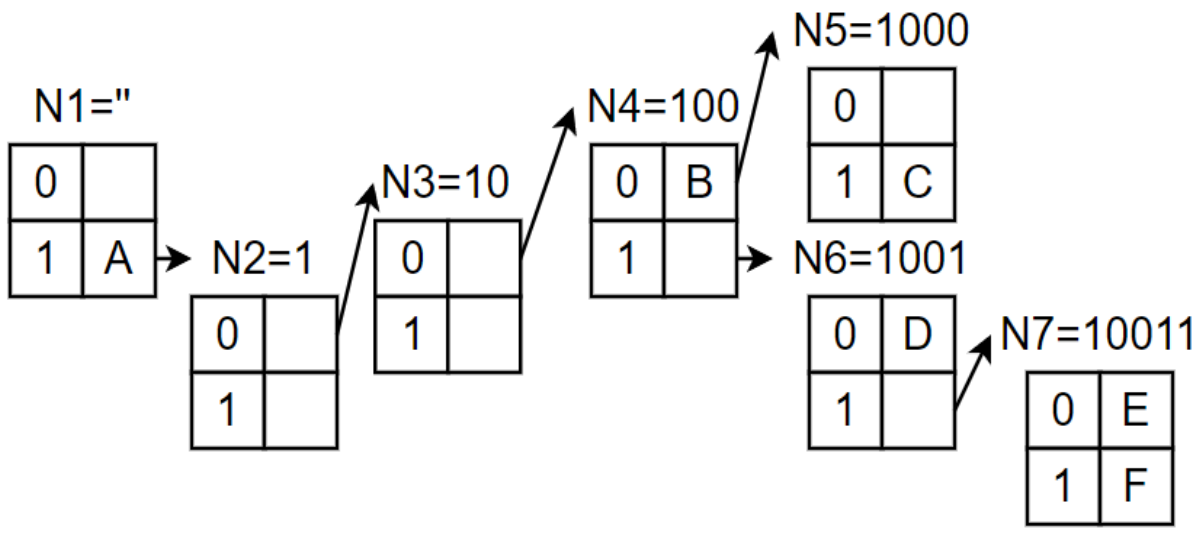

**Figure 2: Unibit Trie for Table 1**

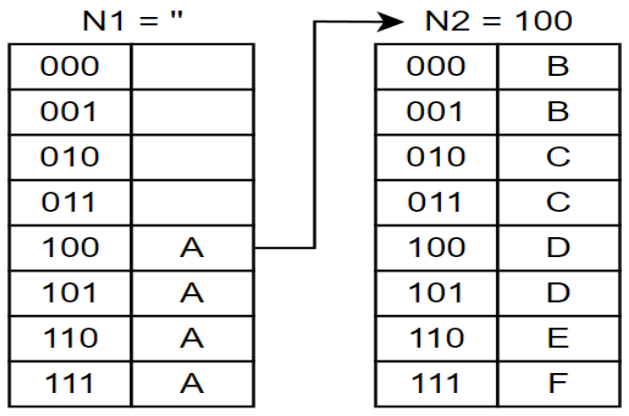

**Figure 3: Multibit Trie for Table 1**

During prefix expansion, prefixes are expanded so their widths align with the strides that are being used for search. For example, in Figure 3, entry 1 in Table 1 was expanded into 4 prefixes (100 to 111). Collisions during expansion must be handled, such as when expanding entries 2 and 3 in Table 1. Both entries produce the expansion 100011 but entry 3 takes priority since it is the longer prefix.

Multibit tries trade memory for time since additional prefixes are generated in order to reduce the number of memory accesses [29]. Prefix expansion is required due to RAM's inability to perform wildcard matches. TCAMs can match wildcards, so prefix expansion is naturally avoided.

## 3 PROBLEM

In this section we reintroduce the problem and provide an example that illustrates how using a single TCAM table can lead to wasted resources

**Problem Statement:** How can we break up a single large TCAM into multiple smaller TCAMs, connected in a tree structure, such that we can scale to larger IP lookup databases that cannot be supported by a single large TCAM?

### 3.1 Inefficiency of a Single TCAM solution

In a single TCAM, inefficiencies arise from prefixes that share sub-prefixes, and very short prefixes. First, a shared sub-prefix is stored multiple times, one for every entry that contains that shared sub-prefix. Figure 4 represents how Table 1 is stored using a single TCAM table (the associated return value is stored in RAM). Note how the shared prefix 100 is stored 5 times, requiring 15 bits for only 3 bits of shared information.

| TCAM | RAM |
|---|---|
| 1***** | A |
| 1000** | B |
| 10001* | C |
| 10010* | D |
| 100110 | E |
| 100111 | F |

**Figure 4: Single TCAM for Table 1**

Although this waste may appear small in this tiny example, backbone routers are nearing 1 million IPv4 entries and 150,000 IPv6 entries [2]. This means that a single initial prefix of size 16 can be repeated a thousand times, wasting tens of thousands of TCAM bits.

A second source of inefficiency is the need to store short prefixes as full entries. Backbone routers typically hold prefixes of various lengths. Storing a 20 bit prefix along side a 56 bit prefix in a single TCAM table of width 88 produces 44 bits of additional waste for the 20 bit prefix. Better tiling can reduce both inefficiencies.

# 4 SOLUTION

In this section we cover the high-level concepts behind MashUp. We then introduce algorithms and methods for building and using the tree.

## 4.1 The Solution

**Solution:** We produce a tree of TCAM tables that allows larger data sets to be supported for a given amount of TCAM by splitting on bits that correspond to lean levels and packing together several trie nodes in order to reduce overhead (Figure 1). Further improvements are realized by swapping certain TCAM tables for SRAM tables.

## 4.2 An Example

Figure 5a depicts the use of a tree of TCAM tables to store the database from Table 1 (the tree is built using a 3-3 stride). By using the tree approach, we can efficiently accommodate the first 3 bits of the address space at the root. This allows the next level of the tree to not be burdened with redundant information and still distinguish entries 2-6 in Table 1.

## 4.3 Building The Tree

The pseudo-code for building a fixed-stride tree using a depth-first approach can be seen in Algorithm 1. Provided a list of strides and a sorted IP prefix database, a tree is built by recursively attempting to add a prefix to a node (a node is equivalent to a table). Case 1 refers to when the prefix ends in the current node, in this case a sanity duplication check is done. If it passes, the entry is added to the current node, and we move onto the next prefix.

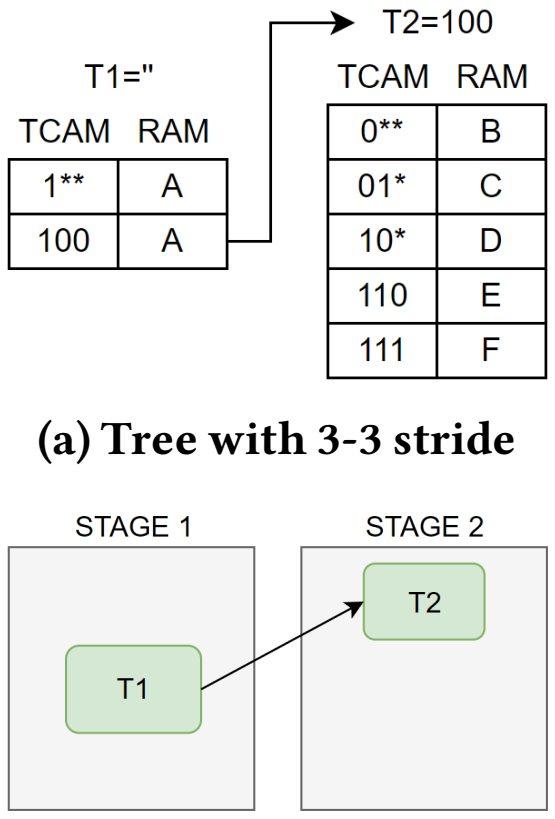

**(a) Tree with 3-3 stride**

**(b) Tree mapped onto a pipeline**

**Figure 5: Tree of TCAMs for Table 1**

Case 2 refers to when the prefix does not end in the current node, in this case a stub key is formed (Note that these stub keys are the pointer overhead mentioned later in §5). We first check if a stub entry exists and if not, we add one and give it the return value of the stub's longest prefix match in the current node. We then check if a child node exists for that stub and if not, we add one. We shorten the prefix by removing the stub portion and repeat the process at the child node. The tree is built after all the prefixes in a database have been inserted.

The strides chosen for the tree can be any size but tables are not built from arbitrary cuts of TCAM, they are implemented by stitching together blocks of a fixed grain size (width by depth). For this reason, instead of looking at the total number of individual bits used, we focus on the total number of blocks required when displaying results in section §6.

The tables in the tree are each mapped onto a set of blocks and the number of blocks required per table is given by

$$x = \left\lceil \frac{TableWidth}{GrainWidth} \right\rceil$$

$$y = \left\lceil \frac{TableDepth}{GrainDepth} \right\rceil$$

$$BlocksforTable = x * y$$

## 4.4 Tagging

The built tree could have several tables that have very few entries. These *barren* tables would lead to significant internal fragmentation in TCAM blocks and could easily exhaust the available TCAM blocks. To rectify this, we prefix entries with a tag so that it is possible for several tables to become a "super-table" to be implemented on the same set of blocks(Figure 1). Every table within a "super-table" has a unique tag so that each table can be searched independently. Tagging is an old idea. For instance, it is used in [23] in the context of "packing TCAMs" for packet classification.

**Algorithm 1** Build-Tree(*Strides*, *SortedDatabase*)

```
Root ← Table w/ width Strides[0]
for all (Prefix, NextHop) pair in SortedDatabase do
    Node ← Root
    Key ← Prefix
    Depth ← 0
    while true do
        if Key.Length ≤ Node.Width then
            {Case 1: When the prefix ends in the current node}
            if Node has entry for Key then
                Report Duplicate Prefix Entry
            else
                Add (Prefix, NextHop) entry to Node
            end if
        else
            {Case 2: When the prefix goes onto the next level}
            Stub ← Key[0 : Node.Width]
            if Node has no entry for Stub then
                Add (Stub, Node.LPMatch(Stub)) entry to Node
            end if
            if Stub.Child = nil then
                Stub.Child ← Table w/ width Strides[Depth+1]
            end if
            Node ← Stub.Child
            Key ← Key[Node.Width : ]
            Depth ← Depth+1
        end if
    end while
end for
```

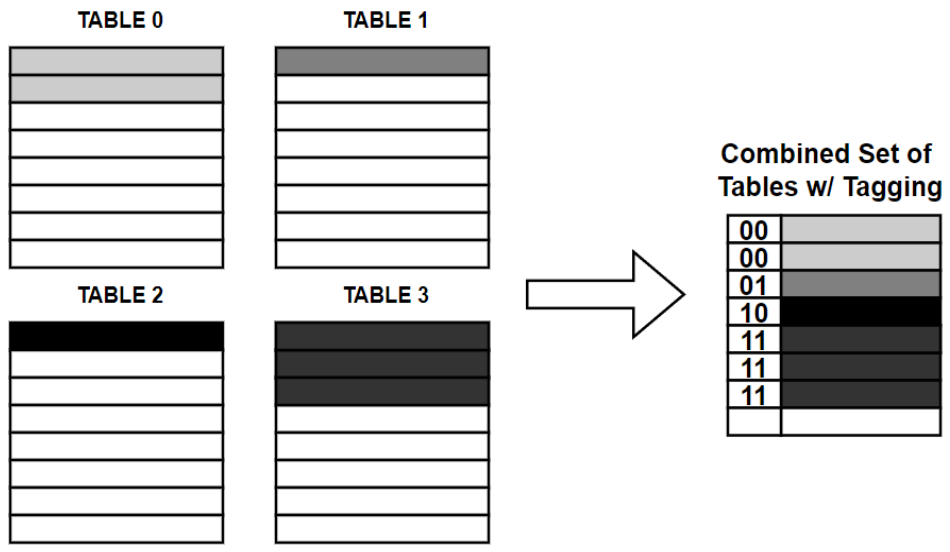

**Figure 6: Tagging Tables to Reduce Waste**

Figure 6 depicts how tag-based aggregation is used to coalesce 4 separate tables, reducing the overall waste produced from empty entries. If table 3 were searched, the search key would be prefixed with 11 and only the entries in the block with that prefix would be candidates since all other entries would fail to match 11.

The cost of using tags is that keys are wider, but this is acceptable because tables are constructed from blocks of a fixed width. In most cases, multiples of this fixed width will exceed the width of the actual key, leaving unused bits, which can be used for tagging. The benefits of tagging are very significant because a tag of width $x$ can potentially save $2^x$ blocks (see §5 and §6).

Coalescing requires choosing *which* tables to pack together into "super-tables". In §5 we show that any algorithm that makes full use of a tag of width $\log_2 D$ ($D$ being grain depth) by packing tables as much as possible suffices to limit the amount of empty entries produced. For example, if the tag width is 8, than any algorithm that packs tables together in sets of 256 will achieve a reasonable bound on empty entries.

For our experimental results in §6, we used an approach that packs tables that have few empty entries with several tables that have many empty entries. This was done with the goal of reducing the worst-case size of a "super-table", aspiring to reduce average power cost. It still, however, meets the constraint to bound empty entries.

## 4.5 Choice of Strides

Algorithm 1 requires a list of strides as input for building the tree, so that begs the question: *What is a good choice of strides for the tree?* In §5 we quantify the amount of overhead that a tree produces for a given stride list. This overhead is linearly proportional to the amount of pointers produced. Fortunately, by building a unibit trie representation of the database, we can easily gather the number of pointers at level $x$ by counting the number of non-leaf nodes at level $x$ in the unibit trie. This is consistent with the intuition of "lean levels" in Figure 1

Algorithm 2 determines which stride lists are acceptable given a set of inputs. We first build the unibit trie. Next, we gather the pointer overhead for every level of the tree (Note that the pointer overhead is a percentage relative to the size of the database). We then note the tag width needed to achieve the bounds mentioned in §5 which is $log_2 D$ where $D$ is the grain depth.

Afterwards, to produce a tree of height $H$ we consider every combination of $H-1$ levels. The final level is always the length of the address space $L$. For every combination, we compute the worst-case overhead as computed in §5. If this overhead lies withing acceptable limits ($B$) we say that this is an acceptable stride list for the tree. Note that multiple acceptable stride lists can exist for a given $B$, and the choice for which to use can depend on various factors, such as power costs or better hybridization.

## 4.6 Optional Hybridization

Although TCAM and SRAM are typically viewed as mutually exclusive for IP lookup, modern products like Tofino-3 include both resources in each stage of their pipelines. It would be a shame to exhaust one resource (say TCAM) while the other resource (say SRAM) is idle and available.

**Algorithm 2** ChooseStrides($Database$,$H$,$W$,$D$,$L$,$B$)

$Uni \leftarrow$ Unibit Trie for $Database$
$LL \leftarrow$ Pointer Overheads for every level of $Uni$
$Tag \leftarrow \lceil log_2 D \rceil$
**for** Every $combination$ in $\binom{Levels}{H-1}$ **do**
  $Overhead \leftarrow 0$
  $Prev \leftarrow 0$
  **for** Every $Lvl$ in $combination$ **do**
    $Overhead \mathrel{+}= (\lceil (Lvl{+}Tag{-}Prev)/W \rceil{+}1)LL[Lvl]$
    $Prev \leftarrow Lvl$
  **end for**
  $Overhead \mathrel{+}= (\lceil (L{+}Tag{-}Prev)/W \rceil{+}1)LL[Lvl]$
  **if** $Overhead < B$ **then**
    $combo$ is an acceptable stride list
  **end if**
**end for**

Recall that a major benefit of applying multibit trie techniques to TCAM is that prefix expansion is avoided, on account of TCAM's wildcard capabilities. However, certain tables in the tree may consist solely of entries that align, or nearly align, with a stride width. Such tables would only require a modest amount of prefix expansion. This is a good exchange because TCAM is significantly less dense than SRAM. Modern products are aware of this, as they typically provide significantly more SRAM than TCAM – e.g., Tofino-3 has three times the number of SRAM pages as TCAM blocks.

This lead us to the idea of *node hybridization*, where certain nodes in the tree are converted to SRAM tables. This conversion is done on a per-table basis and prior to tagging (tagging is mutually exclusive for TCAM and SRAM). We visit each node and calculate what the size of the table would be after prefix expansion. If this expansion is reasonable (less than a given factor $C$) then we mark this table for conversion. The pointer from the parent node can also store a bit that allows the next stage to know whether to do a TCAM or SRAM lookup. This produces additional savings of TCAM blocks, at the cost of added SRAM pages. This mixed resource allocation is a boon on products that provide both TCAM and SRAM; the trade-off can be tuned by adjusting the factor $C$.

We note two reasons why we consider hybridization as optional. First, not all TCAM products readily include both resources, in which case we can only perform MashUp without hybridization. Second, the trees that perform best with hybridization may not be the same trees that perform best without hybridization (an example can be seen in §6). While hybridization is a promising idea, in this paper, we only present preliminary results on its usage; we believe the idea merits further exploration.

## 4.7 Mapping Tables onto Sets of Blocks

It is straightforward to map tables onto blocks of an RMT pipeline. Every "super-table" should be mapped onto a continuous set of blocks in a physical stage. If the number of blocks required is larger than the number of remaining blocks in a physical stage, than this process continues onto an additional physical stage. This does not add much complexity to a search operation as tables that span multiple physical stages need only move onto the next physical stage when a miss is encountered.

The only constraint that must be adhered to is as follows: for any "super-table" ($B$) that is pointed to by another "super-table" ($A$), all the blocks used for $B$ must be from physical stages that occur after all the physical stages used for $A$'s blocks. In other words, this is a sequential dependency stating that a packed set of level $x$ tables can only be implemented on stages that occur after all the stages used for their parents in level $x-1$.

## 4.8 Using The Tree

Once the tree has been built and mapped onto blocks in the pipeline, the process of search is straightforward and similar to searching a multibit trie. The pseudo-code for searching the tree is seen in Algorithm 3.

**Algorithm 3** Search-Tree($Prefix$)

*Table* ← *Root*
*BMP* ← *default*
*StartIndex* ← 0
**while** *Table* ≠ *nil* **do**
  {Step 1: Match}
  *Key* ← *Prefix*[*StartIndex* : *Table.width*]
  *RetBMP, RetTable* ← *Table.Search*(*Key*)
  {Step 2: Action}
  **if** *Table.Match* got a hit **then**
    **if** *RetBMP* ≠ *nil* **then**
      *BMP* ← *RetBMP*
    **end if**
    *StartIndex* ← *StartIndex* + *Table.width*
    *Table* ← *RetTABLE*
  **end if**
**end while**

Given that SDN uses a Match-Action approach, the pseudocode is organized as a match followed by an action based on the returned result. The search begins at the root table and at every table along the search the best matching prefix is potentially updated. Search terminates once we reach a leaf or there is a miss.

Deletion, update, and insertion are nearly identical to search except that upon termination the update algorithm deletes/

changes/inserts the desired entry. These are standard algorithms [29] and not repeated here. However, there is extra consideration when it comes to insertion. [1]

The additional subtlety arises in handling overflow, which is when an entry needs to inserted but the set of blocks does not have a valid location for insertion. In this case, a new block will need to be allocated to the current set of blocks that encounter overflow. If this is not possible, we propose the use of a small overflow TCAM table that acts as a buffer for such entries until the chip can be reconfigured with the updated database.

# 5 THEORETICAL MODEL

We now formulate an upper bound on the savings of MashUp without hybridization, and quantify how close we can come to the maximum savings given a set of constraints.

**Definitions and Assumptions**: Let the Grain Size of each TCAM block be $W$ by $D$, where $W$ is the grain width and $D$ is the grain depth. Let $N$ be the number of prefixes in the database and $Max$ be the Maximum length of a prefix. For IPv4, the database we use [12] has $N \approx 900{,}000$ and $Max = 32$. For IPv6, the database we use [11] has $N \approx 150{,}000$ and $Max = 128$.

However, as we see later, if most of the density (say 99% of prefixes) lies within length $M$, and $M$ is the smallest such number, we will refer to $M$ as the *Maximum Threshold Length*. For our IPv4 databases, $M = 24$ ($99.x\%$ of prefixes are 24 bits or less) and for IPv6, $M = 48$ ($99.x\%$ of prefixes are 48 bits or less). We prefer to use $M$ instead of $Max$ in our estimates below for a fairer comparison because the few prefixes in any database of length $> M$ can be handled by a small overflow TCAM.

We will assume that the unit grains of size $W$ by $D$ can be stacked vertically (*vertical stitching*) or horizontally (*horizontal stitching*). For example, 10 TCAM blocks of $44 \times 512$ can be vertically stitched to make an effective TCAM of $44 \times 5120$, or two TCAM blocks of $44 \times 512$ can be horizontally stitched to make an effective TCAM of $88 \times 512$.

Note that the overhead of horizontal and vertical stitching must be borne by single logical TCAM solutions used today. For example, a single logical TCAM that supports an IPv6 database of 512,000 entries of up to 48-bit prefixes (since most IPv6 prefixes lengths are no more than 48) will require horizontally stitching 2 blocks of 44 by 512 (since 48 does not fit into the grain width of 44 evenly) and vertically stitching 1000 of these horizontally stitched pairs. Since the stitching overhead is very hardware specific, not publicly available (as far as we know), and borne equally for both monolithic TCAMS and MashUp, we ignore it in our analysis below.

In the following analysis, we confine ourselves to using MashUp to produce tries of fixed-stride, but with any number of strides. This includes as a special case using a single stride, wherein our solution reduces to a single logical TCAM.

## 5.1 Upper Bounding Savings

We first prove a lower bound on the number of TCAM bits for *any* solution given TCAM building blocks of $W$ by $D$.

Lemma 5.1. *The number of TCAM bits used by any solution must be at least* $\lceil N/D \rceil * D * W$.

Proof. Since we have $N$ prefixes that must be pointed to by $N$ TCAM entries and these require $\lceil N/D \rceil$ blocks, and each block is $W$ by $D$, the lemma follows. □

Lemma 5.2. *The number of TCAM bits used by a single logical TCAM solution must be at least* $\lceil N/D \rceil * D * W * \lceil (M/W) \rceil$.

Proof. First, observe that we have $N$ prefixes that must be pointed to by $N$ TCAM entries and these require $\lceil N/D \rceil$ blocks. Next, each block is $W$ by $D$, and we need to horizontally stitch together $\lceil (M/W) \rceil$ blocks to “fit” the width $M$ of the prefixes. The lemma follows. □

We put together the last two lemmas to prove:

Theorem 5.3 (Maximum Savings). *The Maximum savings in TCAM bits by using MashUp to produce a fixed-stride trie of TCAMs over a single logical TCAM is a factor of at most* $\lceil (M/W) \rceil$.

Proof. By comparing Lemma 5.2 and Lemma 5.1, we see they differ only by a multiplicative factor of $\lceil (M/W) \rceil$ □

## 5.2 Realizing the Maximum Savings

The previous section showed that the Maximum savings of MashUp (without hybridization) over the single logical TCAM is at most a factor of $\lceil (M/W) \rceil$. In this section, we show that with a reasonable set of constraints, databases that satisfy these constraints (which we argue is most realistic databases), one can get within a few percent of the maximum savings. We now formalize the intuition in Figure 1.

First, consider *lean levels*. Intuitively, one cannot get much benefit from MashUp if the prefixes do not share a comparatively small number (say $N/100$) of sub-prefixes. More concretely, there must be some levels of the corresponding unibit trie that are “lean” in that they have comparatively few pointers that point to the next level. Since pointers represent overhead in a tree not present in a single logical TCAM, picking strides at “lean levels" is crucial to compression efficiency. We formalize this as follows.

*Definition 5.4 (Lean Level).* For any prefix database $P$, let $T$ be the unibit trie representation of $P$. Let $L(b)$ be any number $h$ such that the unibit trie $T$ has at most $Nb/100$ non-leaf nodes at height h.

[1] Although maintaining a TCAM table sorted under insertion and deletion is non-trivial, this is a well studied topic. Methods have been developed to improve insertion times [24], and so we do not address this issue.

Intuitively (Figure 1), this is any level of the unibit trie where the number of non-leaf nodes (nodes with pointers to the next level) is as small as needed (measured by being less than $b\%$ of the total number of database entries). Since these non-leaf nodes represent *pointer entries* that are not present in the database, this is a good place to place a boundary or stride for a trie to reduce overhead. In a general $x$ level trie, we locate $(x-1)$ lean levels that allow us to reach $M$ in $x$ strides. Every level $x$ produces $b_x\%$ overhead and the total overhead is the sum of all $b_x\%$.

Most real IPv6 and IPv4 databases contain a $L(b)$ value for significantly small $b$. For example, the publicly available IPv6 database studied in §6 contains an $L(7)$ at 32 and an $L(0.3)$ at 19. That means that at level 32 and 19, the number of pointer entries to the next level is less than 7% and 0.3%, respectively, of the database size (which has around 150,000 entries).

On the other hand, the pointer overhead for 31 is very high, requiring a $b$ value of 16. This is not surprising as there are many /32 bit prefixes allocated to the RIRs that increase the amount of pointer entries at Level 31 but require no such entries at Level 32. We believe that lean levels for significantly small $b$ are likely to remain for future databases because of hierarchical allocation via registries [5].

We now show our main theorem that states that with appropriate sufficient conditions, a 2-level tree can come within a few percent of the optimal savings described by Theorem 5.3. This is useful when the maximum prefix length $W < M < 2W$, as it is today for IPv6 and hence a 2-level tree suffices.

Theorem 5.5 (Tiling Theorem). *If there exists an $L(b)$ such that $M - L(b) + log_2 D < W$, then there exists a two level version of MashUp that comes close to the optimal (in terms of TCAM bits) predicted by Theorem 5.3. More specifically, it takes $(1+\epsilon)\lceil N/D \rceil * D * W$ TCAM bits to implement IP lookup, where $\epsilon \leq 2b/100$.*

Proof. First, even in this limited theorem, note that the conditions are not very restrictive given today's parameters. In IPv6 for example, we will see that $L(0.3)$ is 19 for public databases where $M = 48$. Using a Tofino-3 width $W = 44$ and depth $D = 512$, we see that $48 - 19 + 9 < 44$, and results in a waste of at most 0.6% from optimal.

The proof proceeds constructively. Intuitively, we build a 2-level trie where the first stride is $L(b)$ bits (a lean length!), and the second stride is $M - L(b)$ bits (remember the maximum prefix length we need to care about is $M$). By the definition of $L(b)$, the maximum number of entries in the root that do not exist in the routing database is $Nb/100$, leading to a wastage of at most $b\%$.

We will also show in the next theorem (Lemma 5.6) that if we allocate $log_2 D$ bits for tagging, we can sufficiently pack enough of the second level trie nodes into sets of blocks to waste at most another $b\%$. This shows that $\epsilon \leq (2b)/100$. Since we need to add a tag "tax" of $log_2 D$ to the remaining width of $M - L(b)$, if $M - L(b) + log_2 D < W$, each trie entry in the second level will fit into a TCAM word, and the theorem is proved by construction. □

It only remains to prove Lemma 5.6.

Lemma 5.6. *If $log_2 D$ tag bits are allocated for any set of $\lceil Nb/100 \rceil$ logical trie nodes, then there exists a simple way to pack these trie nodes into "super-tables" so that at most $Nb/100$ entries are wasted.*

Proof. In some ways, we have to solve a bin packing problem. We may be tempted to solve bin packing into bins of size $D$. While we know that good approximations exist [13] that are within 9/11 of the optimal, the optimal can be very bad. Consider if all the trie nodes are of size $D/2+1$. Then at most 1 node can be fit into one block (of depth $D$) wasting $D/2-1$ entries per block, leading to nearly 50% waste, very far from the small percent we are claiming.

The way out of this quandary is to notice that we can use vertical stitching, a degree of freedom not available to standard bin packing (vertical stitching amounts to aggregating several bins into super bins of any size). If we pack together all the trie nodes and simply vertically stitch several continuous blocks we waste at most 1 block(the final block). Unfortunately, if we have several nodes of size 1, we need many tag bits to prevent aliasing of the bits within each logical trie node. This would be far from the $\log_2 D$ bits of tag tax the theorem wishes to limit itself to.

The way out of this second dilemma is to pack into smaller "super-tables" based on the size of the physical blocks. We simply (for now) aggregate $D$ logical trie nodes into a "super-table" (Figure 1). Each aggregation can waste at most one block at the end. But there are at most $(Nb/100)$ trie nodes (from the previous theorem). Thus there is at most $\lceil (Nb)/(100D) \rceil$ blocks wasted. But each block is of depth $D$ and width $W$. Thus we waste roughly $(NWb)/100$ bits. Since the optimal database size is $(NW)$, this again is at most $b\%$ overhead after packing the trie nodes into TCAM blocks. □

Note, that unlike costs associated with pointers, it is imprudent to reduce packing waste to zero. If every block is completely full, the insertion of any new prefix would cause overflow, needing an additional block to be allocated to a set of logical nodes. Thus, a small bounded amount of empty entries prevents frequent overflow. Further, while we have been highly non-deterministic in our packing strategy while minimizing waste, in practice one may wish to pair large and small tables to reduce the maximum size of a "super-table", thereby reducing worst-case power consumption.

| *Bit to Split On* | *b for L(b)* | *Worst Overhead%* |
|---|---|---|
| 16 | 0.05 | 0.10% |
| 17 | 0.08 | 0.16% |
| 18 | 0.13 | 0.26% |
| 19 | 0.23 | 0.46% |
| 20 | 0.41 | 0.82% |
| ... | ... | ... |
| 31 | 15.70 | 31.40% |
| 32 | 6.88 | 13.76% |
| 33 | 7.68 | 15.36% |
| ... | ... | ... |
| 47 | 32.38 | 64.76% |
| 48 | 0.21 | 0.42% |
| 49 | 0.21 | 0.42% |
| ... | ... | |
| 64 | 0.00 | 0.00% |

**Table 2: Lean Levels for IPv6**

# 6 RESULTS

In this section, we present the resource requirements for implementing IP lookup using MashUp and compare them against a single table implementation. Further, we explore the benefits of reducing the grain size.

## 6.1 Databases

IPv6 results were gathered using the BGP routing table database for AS131072. Only active entries (those in the FIB) were used.[2] This database can be found at bgp.potaroo.net/v6/as2.0/ and is updated periodically. The version that was used to gather results was pulled from the database on January 7th, 2022 and had around 150,000 active entries.

IPv4 results were gathered using the BGP routing table database for AS65000. Only active entries (those in the FIB) were used. This database can be found at bgp.potaroo.net/as2.0/ and is updated periodically. The version that was used to gather results was pulled from the database on November 9th, 2021 and had around 900,000 active entries.

## 6.2 IPv6

The IPv6 address space is 128 bits wide, but only the first 64 bits are used for global routing so stride(s) need only add up to 64. Although 99.36% of prefixes have a length $\leq 48$, all prefixes are included because prefixes beyond 48 do not require special treatment for the single logical TCAM solution and can be implemented as an additional level in the trie when using MashUp, akin to a special purpose TCAM.

Table 2 provides the overhead required for splitting on a certain bit in the trie. The second column represents the relative amount of pointer entries produced at that level when compared to the size of the database $N$. The third column shows the worst case overhead% (relative to $N$) when a tag of $log_2 D = 9$ is used, as referred to in §5.2.

Table 3 shows the results of using MashUp with various strides on the IPv6 database, with a grain size of $44 \times 512$. The bits to split on were various combinations of bits 19, 32, and 48. Bit 19 was chosen because it was the largest bit that allowed us to fit the root into a single block. Bit 32 was chosen because it was a local minimum in overhead, which can be seen in Table 2. Bit 48 represents the length that covers the majority of prefixes in a database (by definition, this is always a good lean level).

The importance of tagging can be seen from the difference between columns 4 and 5. Column 4 shows the number of blocks required before tagging, where several tables contained significant amounts of empty entries. Column 5 shows the number of blocks required after tagging, where several tables were packed into "super-tables" in order to reduce the amount of empty entries.

The final column in Table 3 highlights the factor of improvement for using MashUp (without hybridization) when compared to using a single logical TCAM table (Case 0). These results support the ideas presented in §5.2 because the difference from the optimal $2X$ improvement (Theorem 5.3) lies within the amount predicted from using the bits that were split on.

Although all cases produce good reductions, it should be noted that Cases 3-5 are better than Cases 1-2 because they utilize a small root; this is beneficial since every search is expected to use the root. Among those 3 cases, Case 4 is the optimal choice because it provides the most reduction while only using a tree of height 3 (2 if we exclude the final level that accounts for a minuscule number of exceptionally long prefixes). This is consistent with the theory since Case 4 splits on bits that require the smallest overhead.

Note the product that inspired the use of the our grain size (Tofino-3) also has a limit of 384 TCAM blocks which means that a single table would require too many blocks for IPv6. On the other hand, MashUp would generate a tree of tables that could be implemented within Tofino-3's constraints.

## 6.3 IPv6 with optional Hybridization

§4.6 suggests that further reductions in TCAM can be achieved by converting certain TCAM tables into SRAM tables. Why waste TCAM when there is an abundance of SRAM lying around? For example, Tofino-3 contains 384 TCAM blocks of size $44 \times 512$, but also contains 1280 SRAM pages of size $128 \times 1024$!

Table 4 shows the results of using MashUp with hybridization on the IPv6 database. A conversion factor of 8 was used, which means that if after prefix expansion a SRAM table is less that 8 times the size of the original TCAM table, then that table is implemented in SRAM as opposed to TCAM. The final column estimates the amount of SRAM pages needed to implement the SRAM tables that were produced. Since the Tofino-3 has a page depth of 1024, this estimate is equivalent to $\lceil RAMEntries/1024 \rceil$.

[2] Entries of length 128 were omitted from results because there were only 3.

| Case | Stride(s) | Level(s) | Blocks Pre-Tag | Blocks Post-Tag | Improvement |
|---|---|---|---|---|---|
| 0 | 64 | 64 | 574 | 574 | - |
| 1 | 32-32 | 32-32 | 10214 | 304 | 1.888X |
| 2 | 32-16-16 | 32-48-64 | 10522 | 305 | 1.882X |
| 3 | 19-13-32 | 19-32-64 | 10501 | 305 | 1.882X |
| 4 | 19-29-16 | 19-29-64 | 824 | 289 | 1.986X |
| 5 | 19-13-16-16 | 19-32-48-64 | 10809 | 306 | 1.876X |

**Table 3: IPv6 Results w/ grain size of 44x512**

| Case | Stride(s) | TCAM Blocks | TCAM Improvement | RAM Pages Needed |
|---|---|---|---|---|
| 1 | 32-32 | 240 | 2.392X | 212 |
| 2 | 32-16-16 | 126 | 4.556X | 315 |
| 3 | 19-13-32 | 243 | 2.362X | 70 |
| 3a | 20-12-32 | 242 | 2.372X | 64 |
| 4 | 19-29-16 | 287 | 2.000X | 2 |
| 5 | 19-13-16-16 | 129 | 4.450X | 173 |
| 5a | 20-12-16-16 | 128 | 4.484X | 167 |

**Table 4: IPv6 Results Using Hybridization w/ a conversion factor of 8**

The strides used for Table 4 were based off the strides used for Table 3. We give 3 reasons for this. The first is that hybridization is an optional optimization that is performed after generating the tree and before tagging. In other words, this option does not change the produced tree. The second reason is that hybridization can only improve TCAM requirements since the premise is to swap TCAM tables for SRAM tables. The last reason is that the criteria for building a good hybrid tree differs from the one for building a good TCAM tree, and since hybridization is optional, we do not present methods for finding optimal hybrid trees.

The use of hybridization produces additional improvement for every case seen in Table 3, although the amount varies. Note that the amount of conversion is inversely proportional to a level's width. This is because the amount of entries produced from prefix expansion is proportional to the difference between an entry's length and the table's width.

The added improvement seen in Case 1 was obtained by converting the large root into SRAM. Case 2 mimics this, but further improvement is obtained by later levels with smaller strides. Since a significant amount of the improvement for these 2 cases can be attributed to using a large root, cases 3-5 are of more interest.

Case 3 obtains additional improvement from the first 2 levels of the tree, but fails to do so for the last level (due to the 32-bit stride). Surprisingly, the previously optimal Case 4 fails to make further significant improvement, saving only 2 blocks, the root and 1 of 2 last level blocks. The 29-bit stride in the second level (which covers the majority of the database) is the origin of this failure. This underscores the notion that "pure" TCAM trees and hybrid trees have different criteria for optimization.

Case 5 produces significant additional improvement and is the best example of using hybridization as an optional improvement to MashUp on IPv6. The use of a 4-level tree ensures that the stride for every level is small enough to produce a good amount of conversion.

Case 3a and 5a are small modifications of 3 and 5, and highlight how optimal hybrid trees are not equivalent to optimal TCAM trees. Although the extra savings are small, it should be noted that SRAM requirements have reduced as well. A more thorough exploration of what makes a good hybrid tree is left to future work.

In summary, cases 5 and 5a show how TCAM requirements can improve by nearly a factor of 4.5X at the cost of reasonable amounts of SRAM.

## 6.4 IPv4 with optional Hybridization

According to Theorem 5.3, with a grain width $W$=44, MashUp cannot make any improvements in IPv4 (where $M$=24) using TCAM alone. But if we include optional hybridization, this is no longer the case.

Table 5 shows the results of using MashUp with hybridization on the IPv4 database. A lower conversion factor of 3 was used for IPv4 because the database is much more dense then IPv6. This means that the number of additional entries produced by prefix expansion is smaller. The factor 3 was experimentally chosen to bound the amount of SRAM pages used.

| *Case* | *Stride(s)* | *TCAM Blocks* | *TCAM Improvement* | *RAM Pages Needed* |
|---|---|---|---|---|
| 0 | 32 | 1762 | - | 0 |
| 1 | 16-16 | 1727 | 1.020X | 49 |
| 2 | 12-12-8 | 1353 | 1.302X | 385 |
| 3 | 16-8-8 | 655 | 2.690X | 957 |
| 4 | 10-9-5-8 | 352 | 5.006X | 1154 |
| 5 | 15-4-5-8 | 334 | 5.275X | 1118 |
| 6 | 16-4-4-8 | 235 | 7.498X | 1147 |

**Table 5: IPv4 Using Hybridization w/ conversion factor of 3**

A tag of size 14 was used. Using a tag larger than 9 provides less waste than predicted by §5.2, but its use is justified by the relatively large width of 44.

The strides used for hybridization on IPv4 were experimentally gathered. Case 0 represents the use of a single logical TCAM. Case 1 highlighted that significant gains cannot be gained by using a 2-level tree. Cases 2 and 3 utilize a 3-level tree for better results (Note that case 3 saw a significant increase in SRAM pages to match the significant decrease in TCAM needed). Cases 4-6 employ a 4-level tree to gather the best results by making further significant decreases in TCAM with modest increases in SRAM. Case 6 shines above the rest with an improvement factor of 7.5X.

Recalling that Tofino-3 has 384 TCAM blocks and 1280 SRAM pages, by comparing case 0 to case 6 we see a significant result. With a single logical TCAM solution, the Tofino-3 can only support 22% of the current IPv4 database (we needed 1762 TCAM blocks but only have 384), but using MashUp with hybridization, we are capable of shoehorning the entire database. Furthermore, there is still a significant amount of TCAM left for other purposes, thus answering the question we posed at the start of the paper.

It's clear to see that hybridization produces improvements for TCAM requirements, but a fair question to ask is how an all SRAM implementation of the trie would perform. Table 6 shows a comparison between MashUp with hybridization and a "pure" SRAM trie, based on the amount of resources required (TCAM blocks and SRAM pages).

The trees used were those that performed optimally during hybridization. There are two reasons for this. The first is that if we restrict ourselves to only 4 levels, the corresponding SRAM trees are near optimal. The second is that MashUp is focused on reducing TCAM; in other words we are checking to see whether or not the use of any TCAM is warranted.

For IPv4, the use of hybridization converted 619 SRAM pages into 235 TCAM blocks. At a glance it appears that this amounts to 1 TCAM entry for 2.6 SRAM entries, but recall that SRAM pages have double the depth of TCAM blocks so the actual ratio is 1 TCAM entry for 5.2 SRAM entries. This is a modest ratio and is expected on account of how dense the IPv4 database is.

It is worth noting that the "pure" RAM implementation takes 1766 pages and does not fit within Tofino-3's 1280 SRAM page constraint, just as the "pure" CAM implementation does not fit. Hybridization outperforms both extremes and is particularly attractive for architectures like Tofino 3 that have both SRAM and TCAM.

The results for IPv6 are much more interesting with hybridization converting 152,679 SRAM pages into just 128 TCAM blocks. This amounts to a ratio of 1 TCAM entry for 2400 SRAM entries! This example highlights the issues with prefix expansion, especially in an address space as wide as that of IPv6.

The previous result is a consequence of only limiting ourselves to a 4-level tree. For completeness, we include an additional example where we allow the use of an 8-level tree for IPv6. In this additional example, 80 SRAM pages were converted to 16 TCAM blocks giving us a ratio of 1 TCAM entry for 7.5 SRAM entries. Although this result is less extreme, it does require the use of more levels which naturally favors a RAM implementation (recall unibit tries). It also has the disadvantage of requiring a deeper pipeline. Therefore, the performance of MashUp with hybridization when restricted to a 4-level tree in IPv6 appears to be a significant gain.

The previous results highlight the utility of using MashUp with hybridization. TCAM exhaustion can be attributed to the limited amounts included because of TCAM's high power and area costs. SRAM exhaustion can be attributed to the excessive overhead of prefix expansion because of SRAM's inability to do wildcard matching. MashUp with hybridization pairs both technologies so that they cover each other's weaknesses while allowing each to use the other's strength. Expressed as a slogan, *if prefix expansion is a heavy transaction, then CAM is the plan; on the other hand, if the difference from CAM is less than a gram, we will expand and give it to RAM.*

## 6.5 Choosing Grain Size

As a closing note, we shift gears and put on the hat of a designer of a new chip like Intel's Tofino-4 or Broadcom's Trident-5. What TCAM grain size should the designer use?

| *Version* | *Stride(s)* | *Conversion Factor* | *MashUp TCAM* | *MashUp SRAM* | *Only SRAM* |
|---|---|---|---|---|---|
| IPv4 | 16-4-4-8 | 3 | 235 | 1147 | 1766 |
| IPv6 | 20-12-16-16 | 8 | 128 | 167 | 152,846 |
| IPv6 | 20-12-4-4-4-4-8-8 | 8 | 16 | 263 | 323 |

**Table 6: Hybridization and reductions in SRAM**

| *IPv4* | *IPv6* |
|---|---|
| 39,694,336 | 12,931,072 |

**Table 7: Bits Required w/ Grain Size** $44\times512$

First, we acknowledge that TCAMs are used for various functions besides IP lookup such as exact matching and packet classification. In this section, we only limit ourselves to IP lookup. Second, one must consider the hardware overhead for vertical and horizontal stitching. For now, assume that overhead is inversely proportional to grain size. This refines our question to: *What is the largest grain size for which there can be space savings for both IPv4 and IPv6 using MashUp?*

Now that we are comparing between grain sizes, it would be unfair to use block totals as a comparison, so instead we look at the number of ternary bits. Table 7 shows the total bits required by a single TCAM solution for both IPv4 and IPv6, using a grain of $44\times512$. These numbers serve as our point of comparison.

Table 8 shows the savings of MashUp without hybridization for the IPv4 database as the grain width $W$ varies from 18 to 30 using an optimal 3-level tree with strides 9-15-8.[3]. A grain depth of 256 was used to match the decrease in width. There are trivial savings to be had for simply using a smaller grain width for IPv4 so columns 2 and 3 show the improvement that comes by using a single TCAM with a smaller grain width.

By looking at Table 8, 22 is the largest width with savings that can be attributed to the use of MashUp. Table 9[4] shows the results of applying MashUp with the new grain size of $22\times256$. IPv4 results in a 1.7X improvement without hybridization and a 15X improvement with hybridization. IPv6 results in a 3.5X improvement without hybridization and a 8.5X improvement with hybridization. Thus, ignoring other considerations, a grain size of 22 will provide benefits for both IPv4 and IPv6.

## 7 RELATED WORK

CoolCAMs [32] uses algorithmic techniques to generate a 2-level tree implemented to reduce TCAM power costs. CoolCAMs proposes two techniques. In the first, bit selection, a few bits of each prefix are used to partition the prefix set; each partition is searched by a TCAM. The second idea is a 2-level TCAM architecture like ours in which the first level TCAM points to the second level TCAM. Many extensions to this idea of 2-level trees to reduce power abound, including CoolerCAMs [15] and EaseCAMs [30]. CoolCAMs and CoolerCAMs still require every entry to be fully stored, so there are no reductions in space requirements [15, 32]. A patent from Barefoot networks [3] also uses 2-level trees as in CoolCAM but adds a form of *per-level hybridization*. In the patent, a whole level of a tree can be DRAM or TCAM, unlike our per-node hybridization.

Liu [14] proposes two schemes, pruning and mask extension, to compact the rules of a router table by removing redundant prefixes with the same next hop. They also propose replacing the original set of prefixes by an equivalent set of (more compact) generalized prefixes using logic minimization. Some of these compaction algorithms are extended in EaseCAMs [22]. [17] suggests clever merging of IP databases in virtual routers to reduce TCAM. Both compaction based on next hops and merging of virtual router databases are orthogonal to the ideas in this paper. They can be applied before using MashUp to further reduce TCAM blocks at the cost of more complicated insertion and deletion.

Sahni et al ([16], [20]) augment CoolCAM ideas with wide SRAMs, and store the suffixes of several prefixes in a single wide SRAM word. This reduces both power consumed and total TCAM memory. The idea is similar but not identical to our per node hybridization technique.

Algorithmic techniques are also used to reduce TCAM update costs. One method uses a hybrid of SRAM based pipelines and TCAMs [18]. Another seminal method [24] sorts entries in order of length with prefixes of the same length being arbitrarily sorted. Note that MashUp has fast updates because of its tree structure but the ideas of [24] can be used on a per node basis to improve insertion and deletion costs.

Multiple proposals combine algorithmic techniques and TCAMs for *packet classification*. These include MagicCAM [31], Hybrid CAM [18], and Extended TCAMs [27] but these are not relevant for IP Lookup and require changing the underlying TCAM architecture. TreeCAMs [28] does not change the underlying TCAM architecture, and uses dual decision trees in order to efficiently use TCAM for packet classification while keeping update costs low. [19] uses two levels of classification to reduce power for packet classification.

In summary, none of these approaches use our concepts of lean levels to pick strides, work with existing TCAM architectures like Tofino-3, and provide guidance on fitting the tree to TCAM grain sizes as in Theorem 5.5. Most earlier work focuses on power reduction, fast updates, and/or packet classification. Note that MashUp can reduce power and provide fast updates as a side effect but its major objective is reducing TCAM bits used.

[3] This was chosen using lean levels; the details are omitted for space reasons.
[4] Previous optimal hybrid trees were used; the conversion factor for hybridization was 3 and 8 for IPv4 and IPv6 respectively.

| *Grain Width* | *Single TCAM Total Bits* | *Single TCAM Improvement* | *MashUp Total Bits* | *MashUp Improvement* |
|---|---|---|---|---|
| 18 | 32,467,968 | 1.223X | 32,504,832 | 1.221X |
| 20 | 36,075,520 | 1.100X | 18,119,680 | 2.191X |
| 22 | 39,683,072 | 1.000X | 19,903,488 | 1.994X |
| 24 | 21,706,752 | 1.827X | 21,706,752 | 1.827X |
| 26 | 23,515,648 | 1.687X | 23,515,648 | 1.687X |
| 28 | 25,324,544 | 1.567X | 25,324,544 | 1.567X |
| 30 | 27,133,440 | 1.463X | 27,133,440 | 1.463X |

**Table 8: Effect of Various Grain Widths for IPv4 with Stride 9-15-8**

| *Version* | *Hybridization* | *Total Bits* | *Improvement* |
|---|---|---|---|
| IPv4 | No | 23,153,152 | 1.714X |
| IPv4 | Yes | 2,635,776 | 15.060X |
| IPv6 | No | 3,627,008 | 3.565X |
| IPv6 | Yes | 1,526,272 | 8.472X |

**Table 9: MashUp w/ Grain Size** $22 \times 256$

## 8 CONCLUSIONS

This paper shows how to scale TCAM-based IP lookup in chips with re-configurable pipeline stages containing SRAM and TCAM as in Tofino-3 and possibly even in Broadcom's Trident-4 [7] which also uses a pipeline of reconfigurable tiles. Our journey towards tiling trees began with initial work where we assumed very small grain widths (e.g., 8) and obtained 4X improvements in TCAM bits for IPv4. A meeting with vendors, however, forced us to confront the reality of TCAM grain sizes, but gave us the key tool of tag aggregation. This compelled us to pivot to IPv6.

Scores of simulations, where we found close to 2X improvements for some combination of strides, made us suspect a more general phenomena. This led us to lean levels, and the packing theorem. Finally, given that we needed RAM for pointers, we decided to take a flyer on per node hybridization, which surprised us by its effectiveness.

This is a preliminary study. Real chips will have other limits (*e.g.* on granularity of vertical and horizontal switching) but we are optimistic that the basic ideas can be extended. We have only scratched the surface of node hybridization; in particular extending the theorems to incorporate hybridization will require some way to capture the density of small prefix lengths at certain levels that can cause excessive prefix expansion. Variable stride tries are another degree of freedom worth exploring.

More generally, hybridization and tree tiling suggests the possibility of a theory of complexity for other data structures besides IP lookup trees (for example for packet classification) with a memory hierarchy of SRAM, TCAM, and DRAM, inspired by network memory and packet counters [25] that exploit a memory hierarchy of SRAM and DRAM.